\documentclass[a4paper,11pt]{article}
\pdfoutput=1 

\usepackage{jcappub} 

\usepackage[utf8]{inputenc}
\usepackage[T1]{fontenc} 
\usepackage{graphicx}
\usepackage{xcolor}

\usepackage{cases}
\usepackage{amsmath}
\usepackage{amssymb}
\usepackage{amsfonts}
\usepackage{amssymb}
\usepackage{dcolumn}
\usepackage{bm}
\usepackage{bbm}
\usepackage{array}
\usepackage{subfigure}
\usepackage{hyperref}
\usepackage{wasysym}

\def\be{\begin{equation}}
\def\ee{\end{equation}}
\def\bea{\begin{eqnarray}}
\def\eea{\end{eqnarray}}
\def\ba{\begin{align}}
\def\ea{\end{align}}
\newcommand{\bit}{\begin{itemize}}
	\newcommand{\eit}{\end{itemize}}

\title{\boldmath DHOST Bounce}

\author[a, b, c]{Amara Ilyas}
\author[d, e]{Mian Zhu}
\author[a, b, c]{Yunlong Zheng}
\author[a, b, c,*]{Yi-Fu Cai}
\author[f, g, a,*]{Emmanuel N. Saridakis}

\affiliation[a]{Department of Astronomy, School of Physical Sciences, University of Science and Technology of China, Hefei, Anhui 230026, P.R. China}
\affiliation[b]{CAS Key Laboratory for Researches in Galaxies and Cosmology, University of Science and Technology of China, Hefei, Anhui 230026, P.R. China}
\affiliation[c]{School of Astronomy and Space Science, University of Science and Technology of China, Hefei, Anhui 230026, P.R. China}
\affiliation[d]{Department of Physics, The Hong Kong University of Science and Technology, Clear Water Bay, Hong Kong S.A.R., P.R. China}
\affiliation[e]{HKUST Jockey Club Institute for Advanced Study, The Hong Kong University of Science and Technology, Clear Water Bay, Hong Kong S.A.R., P.R. China}
\affiliation[f]{Department of Physics, National Technical University of Athens, Zografou Campus GR 157 73, Athens, Greece}
\affiliation[g]{National Observatory of Athens, Lofos Nymfon, 11852 Athens, Greece}

\emailAdd{aarks@mail.ustc.edu.cn}
\emailAdd{mzhuan@connect.ust.hk}
\emailAdd{zhyunl@ustc.edu.cn}
\emailAdd{yifucai@ustc.edu.cn}
\emailAdd{msaridak@phys.uoa.gr}

\abstract{We present a new class of nonsingular bounce cosmology free from instabilities, using a single scalar field coupled to gravity within the framework of the Degenerate Higher-Order Scalar-Tensor (DHOST) theories. In this type of scenarios, the gradient instability that widely exists in nonsingular bounce cosmologies in the framework of scalar-tensor and Horndeski/Galileon theories is removed by the effects of new operators introduced by the DHOST, due to the modification that they later bring about to the dispersion relation of perturbations. Hence, our results demonstrate that there is indeed a loophole for this type of bounce scenarios to be free from pathologies when primordial perturbations evolve through the bounce phase, and thus the theoretical {\it no-go} theorem for nonsingular bounce cosmology of Horndeski/Galileon theories can be delicately evaded in DHOST extensions. }

\begin{document}

\maketitle

\section{Introduction}
\label{sec:intro}

With accumulated data, the $\Lambda$-Cold Dark Matter ($\Lambda$CDM) cosmology has become the standard paradigm in describing the evolution of our universe. In this standard paradigm, cosmologists believe that the Universe has undergone a short period of inflationary phase at extremely high energy scales, which can provide a causal mechanism for generating primordial perturbations that are appropriate to seed the formation of the large scale structure (LSS) of the Universe \cite{Mukhanov:1990me}. However, it remain a mystery how the Universe evolves from a big bang singularity which cannot be avoided within the inflationary $\Lambda$CDM cosmology \cite{Borde:1993xh, Borde:2001nh}. To address this conceptual issue, the idea of nonsingular bounce cosmology was investigated, where the universe was assumed to begin the contracting evolution from a low energy state with a large volume, then to experience a nonsingular bouncing phase and afterwards to connect with the observed thermal expansion (see \cite{Novello:2008ra, Lehners:2008vx, Cai:2014bea, Battefeld:2014uga, Brandenberger:2016vhg, Cai:2016hea} for comprehensive reviews).

Bounce cosmology can be regarded as an alternative paradigm to inflation in describing the very early moments of the Universe. In the literature, the ``pre big bang'' cosmology \cite{Gasperini:1992em} considers that the universe could start from an initial state with a very tiny curvature following the scale-factor duality found in string cosmology. Additionally, the scenario of ``ekpyrotic/cyclic'' universe \cite{Khoury:2001wf, Khoury:2001bz} is based on the periodic collision of two membranes in a high-dimensional spacetime. Moreover, the configuration of matter bounce was put forward \cite{Wands:1998yp, Finelli:2001sr}, which can yield almost scale invariant power spectra of primordial perturbations and hence it is of observational interest \cite{Brandenberger:2007by, Cai:2007zv, Cai:2008ed}. A nonsingular bounce solution may be achieved within modified gravity constructions \cite{Nojiri:2006ri, Capozziello:2011et, Cai:2015emx}, such as Lagrange-multiplier modified gravity with limiting curvature \cite{Mukhanov:1991zn, Brandenberger:1993ef, Cai:2010zma, Yoshida:2017swb}, higher-order gravity \cite{Biswas:2005qr, Biswas:2006bs}, $f(R)$ gravity \cite{Nojiri:2014zqa, Pavlovic:2017umo}, $f(T)$ gravity \cite{Cai:2011tc}, $f(R,T)$ gravity \cite{Sahoo:2019qbu}, nonlinear massive gravity \cite{Cai:2012ag}, braneworld scenarios \cite{Shtanov:2002mb, Saridakis:2007cf}, nonrelativistic gravity \cite{Cai:2009in, Saridakis:2009bv}, loop quantum cosmology \cite{Bojowald:2001xe, Cai:2014zga, Cai:2014jla, Odintsov:2015uca}, etc. Furthermore, cosmology of nonsingular bounce may be also investigated by the approach of effective field theory (EFT), namely the introduction of matter sectors that violate the null energy condition (NEC) \cite{Cai:2007qw, 
Cai:2008qw, Cai:2009zp}. 

The realization of a nonsingular bounce, however,  suffers from several conceptual challenges. While the matter bounce cosmology and its generalized scenarios may provide an explanation for the scale invariant power spectrum \cite{Wands:1998yp, Finelli:2001sr, Tsujikawa:2002qc, Cai:2015vzv}, moderate primordial non-Gaussianities may also be derived \cite{Cai:2009fn, Gao:2014hea, Gao:2014eaa}, which indicates an observational {\it no-go} theorem for single-field matter bounce based on a K-essence type Lagrangian, that its parameter space can be constrained due to the severe tension between tensor-to-scalar ratio and non-gaussianities \cite{Quintin:2015rta, Li:2016xjb, Akama:2019qeh}. 
Additionally, bounce models that involve matter fields violating NEC often suffer from the ghost instability as well as gradient instability \cite{Cline:2003gs, Vikman:2004dc, Xia:2007km, Easson:2016klq}.
Thus, one needs to delicately construct models, namely by using the Horndeski/Generalized Galileon type theories \cite{Nicolis:2008in, Deffayet:2011gz, Horndeski:1974wa, Kobayashi:2011nu}, in order for a ghost-free nonsingular bounce may to be achieved \cite{Qiu:2011cy, Easson:2011zy}. Finally, on top of these, the homogeneous and isotropic background of a bouncing solution is unstable to the development of radiation and anisotropic stress \cite{Karouby:2010wt, Karouby:2011wj, Bhattacharya:2013ut, Cai:2013vm}, which is known as the famous BKL instability \cite{Belinsky:1970ew}.
Having these in mind, an interesting question has been raised in \cite{Cai:2012va}, namely whether a healthy version of nonsingular bounce may be achieved from the perspective of EFT. In their work it was shown that if one combines an era of ekpyrotic contraction with a nonsingular bounce by introducing a scalar field with a Horndeski-type non-standard kinetic term and a negative exponential potential, then most of the conceptual issues can be well addressed and primordial perturbations can evolve through the bouncing phase smoothly \cite{Osipov:2013ssa, Qiu:2013eoa, Battarra:2014tga, Qiu:2015nha, Banerjee:2016hom, Ijjas:2016tpn, Ijjas:2016vtq, Ijjas:2017pei, Saridakis:2018fth}, except that the sound speed squared ($c_s^2$) of primordial scalar perturbations would become negative around the bounce point, triggering the issue of gradient instability. 

The gradient instability issue was later examined comprehensively in \cite{Libanov:2016kfc, Kobayashi:2016xpl, Kolevatov:2016ppi, Akama:2017jsa} where a theoretical {\it no-go} theorem was presented that, for a generic Horndeski type theory as well as some extended versions, nonsingular cosmological evolutions with flat spatial sections suffer in general from gradient instabilities or pathologies in the tensor sector. To evade this theorem, a nonsingular bounce free of ghost and gradient instabilities may be realized in Horndeski cosmology by breaking certain assumptions applied in the theorem \cite{Banerjee:2018svi} or in cuscuton bounce by freezing the scalar degree of freedom \cite{Boruah:2018pvq, Quintin:2019orx}. Otherwise, based on the correspondence between the EFT formalism and Horndeski/Generalized Galileon theories made in \cite{Gleyzes:2013ooa, Cheung:2007st, Weinberg:2008hq}, the gradient instability issue can be avoided by modifying the dispersion relation for perturbations with the help of certain operators \cite{Cai:2016thi, Creminelli:2016zwa, Cai:2017tku, Cai:2017dyi, Cai:2017dxl}. In particular, recent investigations have shown that the EFT operators $^{(3)}R \delta g_{00}$ as well as $^{(3)}R \delta K$ (here$^{(3)}R$ and $K$ are the intrinsic and extrinsic curvature in the ADM formalism) could be employed to keep $c_s^2$ positive. Remarkably, these needed operators can naturally arise in covariant forms within the framework of the Degenerate Higher-Order Scalar-Tensor (DHOST) theories \cite{Langlois:2015cwa, Langlois:2017mdk}, and in particular the coupling of the scalar field and gravity could be that of the DHOST form. Hence, in the present work we propose a new cosmological scenario of nonsingular bounce, starting from an explicitly covariant Lagrangian combining DHOST terms with the original Lagrangian developed in \cite{Cai:2012va}. In this way all the above issues are successfully evaded, and moreover, as we will see, the gradient instability problem can be terminated as well due to the DHOST term.

The article is organized as follows. In Section \ref{sec:dhostbounce} we briefly present the motivation for developing a model of DHOST bounce, and then provide the generic action. Then, the physics of DHOST bounce will be derived in Section \ref{sec:GeneralAnalysis}; a realization of DHOST bounce will be presented in Section \ref{sec:realization}, and its detailed background evolution and perturbation analysis of the model under construction can be found in Section \ref{sec:BackgroundEvolution} and \ref{sec:CosmologicalPerturbation} respectively. In Section \ref{sec:example}, we provide two type detailed examples in Section \ref{sec:fX} and \ref{sec:fphi} respectively, to illustrate that our model is free of gradient and ghost instabilities. One potential problem, the superluminality issue is discussed in Section \ref{sec:superluminality}. Finally, Section \ref{conclusion} is devoted to a summary of our results with relevant discussions. Lastly, we provide an introduction to DHOST theory and some detailed calculations in the Appendix for the convenience of the reader.

Throughout the work we define the reduced Planck mass by $M_p \equiv 1/\sqrt{8\pi G} = 1$, where G is the Newton's gravitational constant, and we consider the signature of the metric as $(+,-,-,-)$. The dot symbol represents differentiation with respect to the cosmic time $t$: $\dot{\phi} \equiv d\phi/dt$, and a comma in the subscript denotes a normal derivative: $\phi_{,\mu} \equiv \partial_{\mu} \phi$. Additionally, at the bounce point we set the scale factor as $a_B = 1$ and the cosmic time as $t_B=0$. Finally, we use the subscripts $f_{X}$ and $f_{\phi}$ to denote $f_{X} \equiv {\partial f}/{\partial X}$ and $f_{\phi}\equiv {\partial f}/{\partial \phi}$, respectively.

\section{Generic Action of DHOST bounce} \label{sec:dhostbounce}

The DHOST theories allow for the presence of second-order derivatives of a scalar field ${\phi}$, i.e. of ${\nabla}_{\mu}{\nabla}_{\nu} \phi $, in the Lagrangian, as in Horndeski theories. However, in contrast to the latter, which are restricted to Lagrangians leading to second-order Euler-Lagrange equations (for both the metric and the scalar field), DHOST theories actually allow for higher order Euler-Lagrange equations but are required to contain only one propagating scalar degree of freedom  and two tensor degrees of freedom coming from graviton polarizations in order to avoid the Ostrogradski instabilities \cite{Ostrogradsky:1850fid, Woodard:2015zca}.

It was then acknowledged that the crucial property shared by these models is the degeneracy of their Lagrangian, which guarantees the absence of a potentially disastrous extra degree of freedom \cite{Langlois:2015cwa}. The absence of an extra degree of freedom was confirmed, for beyond Horndeski theories by their relation to Horndeski theories via field redefinition \cite{Gleyzes:2014qga, Crisostomi:2016tcp}, as well as by a Hamiltonian analysis for a particular quadratic case \cite{Deffayet:2015qwa}, and for general quadratic DHOST theories by a general Hamiltonian analysis \cite{Langlois:2015skt}. The absence of extra degree of freedom in these type of theories motivates one to use quadratic DHOST theories for finding the compatible solution of geodesically complete cosmologies and to avoid the various conceptual pathologies \cite{Cai:2012va}. It is shown that beyond Horndeski theories can be free from Ostrogradsky instabilities only if the Hessian matrix is degenerate, and it is this kind of theories that are refered as DHOST \cite{Langlois:2015skt, BenAchour:2016fzp, Langlois:2017mxy}. This fact lies behind the motivation of the present study on a possibly healthy bounce model based on the DHOST theory.

DHOST theories are defined to be the maximal set of scalar-tensor theories which generalize theories of Horndeski type in four dimensional spacetime by including all possible terms with at most three powers of second derivatives of the scalar field, while propagating at most three degrees of freedom. The most general DHOST action involving up to cubic powers of second derivative of the scalar field can be written as 
\begin{align}
\label{dhost32}
 S[g,\phi] = \int d^4x \sqrt{-g} \Big[ & f_2(\phi, X) R +  C^{\mu \nu \rho \delta}_{(2)} \phi_{\mu \nu} \phi_{\rho \delta} + f_3(\phi, X) G_{\mu \nu} \phi^{\mu \nu} \nonumber \\ 
 & +C^{\mu \nu \rho \delta \alpha \beta}_{(3)} \phi_{\mu \nu} \phi_{\rho \delta} \phi_{\alpha \beta} \Big] ~.
\end{align} 
The tensors $C_{(2)}$ and $C_{(3)}$ represent the most general tensors constructed with the metric $g_{\mu \nu}$ as well as the first derivative of the scalar field  which is denoted as $\phi_{\mu} \equiv \nabla_{\mu} \phi$. The symbol $\phi_{\mu\nu}$ denotes the $\phi_{\mu\nu} \equiv \nabla_{\mu} \nabla_{\nu} \phi$, while the canonical kinetic term $X$ becomes $X \equiv \frac{1}{2} \nabla^{\mu} \phi \nabla_{\mu} \phi$. Exploiting the symmetry in $C_{(2)}$ and $C_{(3)}$, one can reformulate Equation \eqref{dhost32} to be:
\begin{equation}
 C^{\mu \nu \rho \delta}_{(2)} \phi_{\mu \nu} \phi_{\rho \delta} +  C^{\mu \nu \rho \delta \alpha \beta}_{(3)} \phi_{\mu \nu} \phi_{\rho \delta} \phi_{\alpha \beta} = \sum_{i=1}^5 a_i L_i^{(2)} + \sum_{j=1}^{10} b_j L_j^{(3)} ~,
\label{extraterms}
\end{equation}
where $a_i$'s and $b_i$'s depend only on $\phi$ and $X$.  For the concrete construction of a bounce model, we only keep $a_i$ terms which are associated with quadratic power Lagrangians $L_i^{(2)}$, and we assume all $b_j$ terms to vanish ($b_j = 0$) in order to eliminate the effects of cubic power Lagrangians $L_j^{(3)}$. Accordingly, the Lagrangians related to $a_i$ terms are listed as:
\begin{align} 
\label{eq:L_i^2}
 & L_1^{(2)} = \phi_{\mu \nu} \phi^{\mu \nu} ~,~ L_2^{(2)} = (\Box \phi)^2 ~,~ L_3^{(2)} =  (\Box \phi)\phi^{\mu}\phi_{\mu \nu} \phi^{\nu} ~, \nonumber\\ 
 & L_4^{(2)} = \phi_{\mu}\phi^{\mu \rho}\phi_{\rho \nu}\phi^{\nu} ~,~ L_5^{(2)} = (\phi^{\mu}\phi_{\mu \nu}\phi^{\nu})^2 ~.
\end{align}

In order to see possible types of DHOST theory of pure quadratic order, we refer to the descriptions presented in Appendix \ref{appen:dhost}. For our purpose of developing a covariant form of the action, we rewrite the theory as
\begin{align} \label{eq:bounceS}
 S = \int d^4x \sqrt{-g} \Big[ & - \frac{1 + f(\phi, X)}{2} R + K(\phi,X) + Q(\phi,X) \Box \phi \nonumber \\ 
 & + \frac{J(\phi,X)}{2} \big( L_1^{(2)} - L_2^{(2)} \big) + \frac{A(\phi,X)}{2} \big( L_4^{(2)} - L_3^{(2)} \big) \Big] ~.
\end{align}
In the action \eqref{eq:bounceS} the first term $-{R}/{2}$ corresponds to the standard Einstein-Hilbert action, 
and $K + Q \Box \phi$ is a type of the Galileon action which has been introduced to the study of cosmology such as in \cite{Qiu:2011cy, Easson:2011zy, Deffayet:2010qz, Kobayashi:2010cm, Cai:2012va}. 
For the rest terms we make the choice
\begin{align} \label{eq:JAforms}
 J(\phi, X) \equiv -\frac{f}{2X} ~,~ A(\phi, X) \equiv \frac{f-2Xf_X}{2X^2} ~, 
\end{align}
while $L_i^{(2)}$ have been provided in Equation \eqref{eq:L_i^2}. We mention that the action in \eqref{eq:bounceS} is originally derived from the $^{(2)}N$\textendash $II$ type quadratic DHOST theory by taking 
\begin{align}
 f_2 = \frac{f}{2} ~,~ a_1 = -a_2 = -\frac{f}{4X} = \frac{J}{2} ~,~ a_4 = -a_3 = \frac{A}{2} = \frac{f-2Xf_X}{4X^2} ~,~ a_5 = 0 ~, 
\end{align}
and is free from Ostrogradsky instabilities.

Following the EFT dictionary developed in \cite{Cai:2017dyi}, one expects that the term
\begin{align}
\label{SR3g00}
 S_{R^{(3)} \delta g^{00}} =  \int d^4x & \sqrt{-g}  \bigg\{ - \frac{f}{2} R + X \int f_{\phi \phi} d\ln X - \Big( f_{\phi} + \int \frac{f_{\phi}}{2} d\ln X \Big) \Box \phi  \\
 & - \frac{f}{4X} \big[ \phi_{\mu\nu} \phi^{\mu\nu} \!- \!(\Box\phi)^2 \big] - \frac{f -2X f_{X}}{4 X^2} \Big[ \phi^{\mu} \phi_{\mu\rho} \phi^{\rho\nu} \phi_{\nu} - (\Box\phi) \phi^{\mu} \phi^{\nu} \phi_{\mu\nu} \Big] \bigg\} ~, \nonumber
\end{align}
is able to solve the gradient instability. Note that the first line of Equation \eqref{SR3g00} comes from the Horndeski form with $K + Q \Box \phi$, while the second line is exactly that of the $^{(2)}N$\textendash $II$ type DHOST Lagrangian. Hence, Equation \eqref{SR3g00} is included in the generic action \eqref{eq:bounceS}. Accordingly, we expect that our model is able to give a nonsingular bouncing solution without gradient instabilities. We shall confirm this assertion in following analyses.

\section{General analysis of DHOST bounce}
\label{sec:GeneralAnalysis}
	
\subsection{A realization of bounce cosmology}
\label{sec:realization}

In this section we consider a scenario which allows for the bounce realization. Inspired by the nonsingular bounce cosmology developed in \cite{Cai:2012va}, we consider the function $ K(\phi,X)$ of action \eqref{eq:bounceS} to have the form
\begin{align}
 K(\phi,X) = [1-g(\phi)]X + \beta X^2 - V(\phi) ~,~~ Q(\phi,X) = \gamma X ~, 
\end{align}
with $\beta$ and $\gamma$ being the model parameters. The potential $V$ and the coupling function $g$ depend on $\phi$ only, with their forms being given respectively by 
\begin{eqnarray}
 V(\phi) = - \frac{2V_0}{e^{-\sqrt{\frac{2}{q}}\phi}+e^{b_V\sqrt{\frac{2}{q}}\phi}}  ~,~ ~ g(\phi) = - \frac{2g_0}{e^{-\sqrt{\frac{2}{p}}\phi}+ e^{b_g\sqrt{\frac{2}{p}}\phi}} ~.
\end{eqnarray}
Moreover, as mentioned in \eqref{eq:bounceS}, the function $f$ is an arbitrary function of $\phi$ and $X$. Thus, different choices of $f$-forms yield different models of the DHOST bounce cosmology.

We apply the ADM decomposition by writing the background metric as
\begin{eqnarray}
 ds^2 = N^2 dt^2 - h_{ij} (dx^i + N^i dt) (dx^j + N^j dt) ~,
\end{eqnarray}
and the DHOST action \eqref{eq:bounceS} 
\begin{equation}\nonumber
 S=\int dt d^3x ~ N\sqrt{h} \mathcal{L}_{ADM} ~,
\end{equation}
where $\mathcal{L}_{ADM}$ is given by 
\begin{eqnarray}
\mathcal{L}_{ADM} = \frac{1+f}{2} \mathcal{R} + \frac{1}{2} \big( \mathcal{K}_{ij}\mathcal{K}^{ij}-\mathcal{K}^{2} \big) + K + \big( -f_{\phi}+Q \big) B \mathcal{K} + Q W ~, \nonumber
\end{eqnarray}
with 
\begin{align}
 B = \phi_{,\mu} n^\mu = \frac{\dot{\phi}}{N} ~,~ ~ W = B_{,\mu} n^\mu = \frac{1}{N} \big( \dot{B} + B \frac{N^{i} \partial_{i} N}{N}\big) ~.\nonumber
\end{align}
Finally, after some algebra we can derive a simplified background action within a spatially flat Friedmann-Lema\^{i}tre-Robertson-Walker (FLRW) geometry, namely
\begin{align}\label{eq:S0}
 S_0=&\int dt dx^3 ~ N a^3 \mathcal{L}_0 ~,
\end{align}
with 
\begin{align}
\label{fu16}
 \mathcal{L}_0 =& \frac{3}{a N} \frac{d}{dt} \Big( \frac{\dot{a}}{N} \Big) (1+f) + \frac{3 \dot{a} }{N^3 a} \dot{\phi} \frac{d}{dt} \Big( \frac{\dot{\phi}}{N} \Big) f_X + \frac{3 \dot{a}^2}{N^2 a^2} (1+2f) \nonumber \\
 & + \frac{3 \gamma \dot{a} \dot{\phi}^3}{2a N^4} + (1-g) \frac{\dot{\phi}^2}{2N^2} +  \frac{\beta\dot{\phi}^4}{4 N^4} + \frac{\gamma 
\dot{\phi}^2}{2N^3} \frac{d}{dt} \Big( \frac{\dot{\phi}}{N} \Big) -  V ~.
\end{align}

\subsection{Background Dynamics} \label{sec:BackgroundEvolution}
 
Let us now extract the field equations of the aforementioned action. These equations can be calculated with the help of the stress energy tensor or alternatively through variation of the background action \eqref{eq:S0} with respect to $N$, $a$ and $\phi$, respectively. In this article we follow the second method, yielding
\begin{align}
\label{E1}
 & \frac{1}{6} (1-g) \dot{\phi}^2 + \frac{1}{4} \beta \dot{\phi}^4 + \gamma H \dot{\phi}^3 + \frac{V}{3} - H^2 = f_{a1} ~, \\ 
\label{E2}
 & \frac{1}{2}(1-g)\dot{\phi}^2 + \frac{1}{4} \beta \dot{\phi}^4 - \gamma \dot{\phi}^2\ddot{\phi} - V(\phi)+ 3H^2+2\dot{H} = f_{a2} ~,\\
\label{E3}
 & \mathcal{\Tilde{P}}\ddot{\phi} + \mathcal{\Tilde{D}} \dot{\phi} + V_{\phi} = f_{a3} ~,
\end{align}
where $N=1$ has been imposed in the final result. The parameters $f_{a1}$, $f_{a2}$, $f_{a3}$ quantify the effect of the DHOST terms at the background equations and are given as 
\begin{align} 
\label{eq:fas}
 f_{a1} & \equiv H \dot{\phi } f_{\phi } + H \dot{\phi}^3 f_{\phi X} ~, \nonumber \\
 f_{a2} & \equiv -\big[ f_{\phi \phi }\dot{\phi }^2 + (f_{\phi } + \dot{\phi }^2 f_{\phi X}) \ddot{\phi} \big] ~, \nonumber \\
 f_{a3} & \equiv 3 (\dot{H} +3 H^2)(f_\phi +f_{\phi \text{X}} \dot{\phi}^2) + 3 H \dot{\phi}^3 (f_{\phi\phi \text{X}} + f_{\phi \text{XX}} \ddot{\phi}) ~,
\end{align}
while  
\begin{align}
 \mathcal{\Tilde{P}} &= 1 -g + 6 \gamma  H \dot{\phi} + 3 \beta \dot{\phi}^2 ~, \nonumber \\
 \mathcal{\Tilde{D}} &= 3 (1-g) H + \big( 9 \gamma  H^2 + 3 \gamma \dot{H} - \frac{g_{\phi}}{2} \big) \dot{\phi} + 3 \beta H \dot{\phi}^2 ~. \nonumber
\end{align}

Equation \eqref{E3} is the dynamical equation of the scalar field $\phi$. We follow the convention and use Equations \eqref{E1}, \eqref{E2} to eliminate the $\dot{H}$ inside and then acquire
\begin{eqnarray}\label{E3b}
 \mathcal{P} \ddot{\phi} + \mathcal{D} \dot{\phi} + V_{\phi} = f_{a3} - \frac{3}{2} \gamma \dot{\phi}^2 (3f_{a1}+f_{a2}) ~, 
\end{eqnarray}
where the functions $\mathcal{P}$ and $\mathcal{D}$ read as
\begin{align}
 & \mathcal{P} = 1 -g +6 \gamma H \dot{\phi} +3 \beta \dot{\phi}^2 + \frac{3}{2} \gamma^2 \dot{\phi}^4 ~, \nonumber \\
 & \mathcal{D} = 3(1-g) H + \big( 9 \gamma H^2 -\frac{g_{\phi}}{2} \big) \dot{\phi} +3 \beta H \dot{\phi}^2 - \frac{3}{2}(1-g) \gamma \dot{\phi}^{3} - \frac{9 \gamma^{2} H \dot{\phi}^{4}}{2} - \frac{3 \beta \gamma \dot{\phi}^{5}}{2} ~.
\end{align}

In the detailed examples of the model building which shall be discussed in the next section, we shall use a set of parameters as follows,
\begin{align}
\label{eq:bgpara}
& V_0 = 10^{-8} ~,~~ g_0 = 1.1 ~,~~ \beta = 5 ~,~~ \gamma = 3 \times 10^{-3} ~, \nonumber \\
& b_V = 100 ~,~~ b_g = 0.5 ~,~~ p = 0.01 ~,~~ q = 0.1 ~,
\end{align}
where all parameters are expressed in units of the reduced Planck mass in the rest of the paper. 
As the background equations depend on the DHOST term, we need to fix the explicit forms of the DHOST term to acquire the numerical evolution. For simplicity, we only consider the cases of $f=f(X)$ and $f=f(\phi)$ in the next section. Before going to these cases, we would like to take a look at the generic discussion on cosmological perturbation in the following subsection.

\subsection{Cosmological perturbations}\label{sec:CosmologicalPerturbation}

This section is devoted to the study of cosmological perturbations within the scenario at hand. For the DHOST bounce model there are three physical degrees of freedom, namely one scalar and two tensor modes.

We start with tensor perturbations first. The generic property of tensor perturbations in the scalar-tensor theory is discussed in \cite{Gao:2019liu}, and the quadratic action for tensor modes in the FLRW background takes the generic form as
\begin{equation}
\label{eq:S2tgeneral}
 S_{2,T(General)} = \int dt d^3x \frac{a^3}{2} \left( \dot{\gamma}_{ij} \hat{\mathcal{G}}^{ij,kl}\dot{\gamma}_{kl} - \gamma_{ij} \hat{\mathcal{W}}^{ij,kl} \gamma_{kl} \right) ~,
\end{equation}
in which $\gamma_{ij}$ are tensor perturbations and $\hat{\mathcal{G}}$,$\hat{\mathcal{W}}$ are determined by the theory. Then, we can substitute our action \eqref{eq:bounceS} into \eqref{eq:S2tgeneral} and get 
\begin{equation}
\label{eq:S2tDHOST}
 S_{2,T(DHOST)} = \int d\tau d^3x \frac{a^2}{8}\left[ \gamma_{ij}^{\prime 2} - (1+f)(\nabla_k \gamma_{ij})^2 \right] ~,
\end{equation}
where $\tau$ is the conformal time defined by $d\tau = dt/a$, and a $\prime$ represents the differentiation with respect to $\tau$. 
One can straightforwardly see from Equation \eqref{eq:S2tDHOST} that the ghost problem is absent in our case. The propagation speed of tensor modes is expressed by $c_T^2 = 1+f$. In order to ensure the tensor modes of our model are free from gradient instability, the condition of $1+f > 0$ is required to be satisfied. Additionally, since $c_T^2 \simeq 1$ today, we expect $f \simeq 0$ in the late universe. Finally, as the scope of the present study is upon the construction of a stable bounce model, we leave the discussion on primordial gravitational waves as well as confrontation with observations in a future project.

After that, we come to scalar perturbations. We work in unitary gauge for convenience and we present the lengthy calculation of the detailed perturbation expansions in Appendix \ref{appen:perturbations}. Here we summarize the finally obtained reduced quadratic action for the curvature perturbation $\zeta$, which acquires the form:
\begin{align}
 S = \int d \tau d^{3} x \frac{z_s^{2}}{2}  \Big[ \zeta^{\prime 2}-c_{s}^{2}\left(\partial_{i} \zeta \right)^{2} \Big] ~,
\end{align}
where
\begin{align}
\label{eq:zs2cs2}
 & \frac{z_s^2}{2a^2} =  3 + 2 \Big[ (Q_{X} -f_{\phi X}) \dot{\phi}^{3} - f_{\phi} \dot{\phi} -2 H \Big]^{-2} ~ \Big[ \dot{\phi}^{2} \big( K_{X} - 2 Q_{\phi} \big) + \dot{\phi}^{4} \big( K_{X X}-Q_{\phi X} \big) - 6 H^{2} \nonumber \\
 & \ \ \ \ \ \ \ \ \ \ \ \ \ \ - 3 H \dot{\phi} \Big( 2 f_{\phi}+\dot{\phi}^{4} f_{\phi X X} + 5 \dot{\phi}^{2} f_{\phi X} \Big) +12 H Q_{X} \dot{\phi}^{3} + 3 H \dot{\phi}^{5} Q_{X X} \Big] ~, \nonumber \\ 
& (-\frac{z_s^2}{2a^2})c_s^2 = 1 + f + \frac{2}{a} \frac{d}{dt}  \bigg[ \frac{a \big( f_{X} \dot{\phi}^{2} - f - 1 \big) }{ 2 H - ( Q_{X} - f_{\phi X}) \dot{\phi}^{3} + f_{\phi} \dot{\phi} } \bigg] ~. \nonumber
\end{align}

We can examine the absence of ghost and gradient instabilities by determining the positivity of the parameters  $z_s^2$ and the sound speed squared $c_s^2$ respectively, under the specific investigated models. In the rest of the paper we will numerically examine the posibility of $z_s^2$ and $c_s^2$ in specific models.

\section{Specific examples for stable DHOST bounces} \label{sec:example}

In this subsection, we consider several concrete examples that are obtained by choosing different forms of the function $f$. We deal with a simple case $f=f(X)$ in which the DHOST correction does not affect the background behaviour of the Horndeski bounce. Then, we consider another case $f=f(\phi)$ where the background dynamics can be dramatically affected. For both cases, the DHOST term is found to be able to cure the gradient instability around the bounce phase, which is the main point of the present work.

\subsection{Case 1: $f=f(X)$}
\label{sec:fX}

\begin{figure}[hb]
\begin{minipage}[t]{0.45\linewidth}
\centering
\includegraphics[scale=0.31]{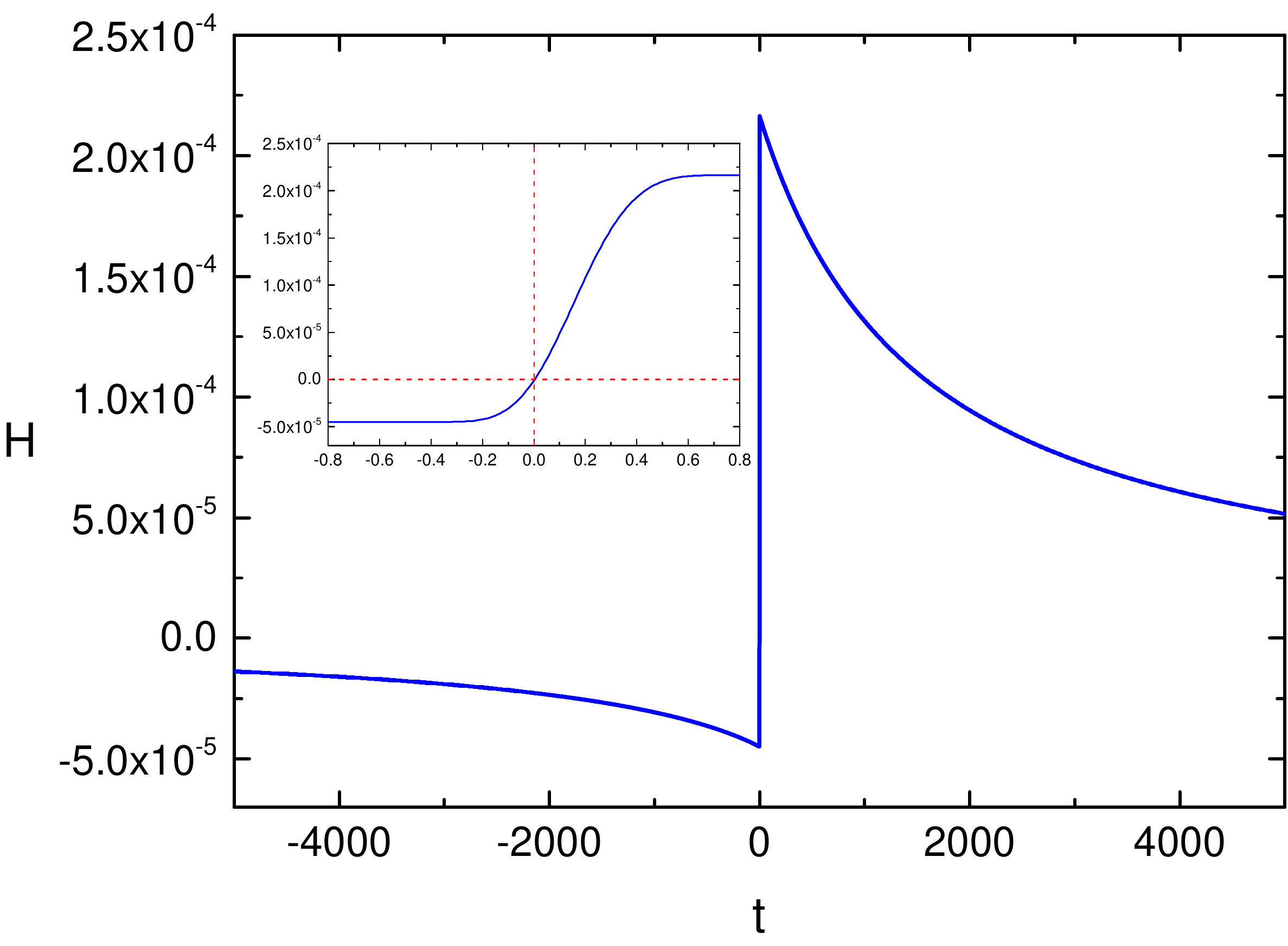}
\end{minipage}
\hspace{.3in}
\begin{minipage}[t]{0.45\linewidth}
\centering
\includegraphics[scale=0.3]{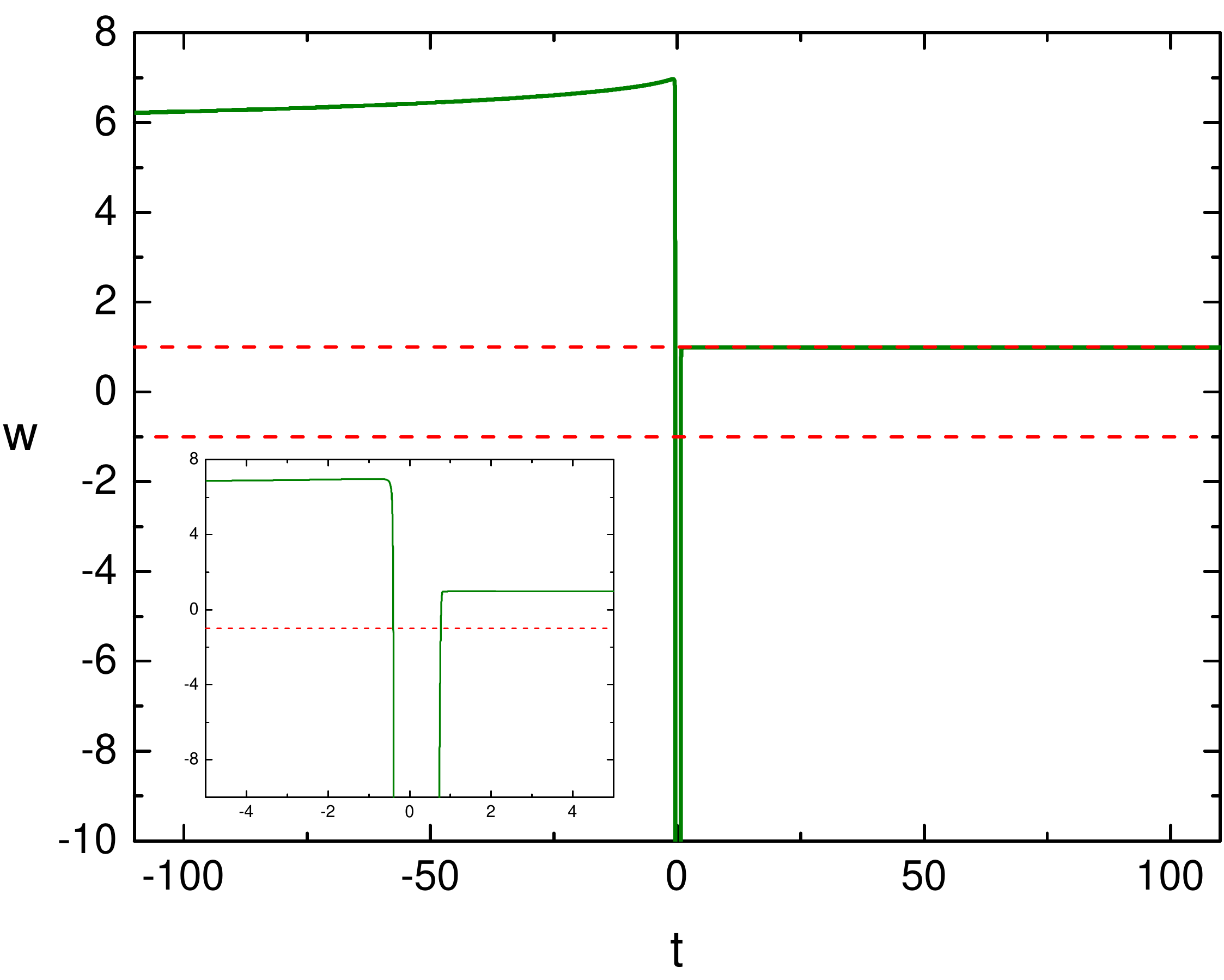} 
\end{minipage}	
\caption{Numerical evolution of the Hubble parameter $H$ and the equation-of-state parameter $w$ as functions of cosmic time when $f=f(X)$. The main plot shows that a nonsingular bounce occurs, and that the time scale of the bounce is short (which is referred as \textit{fast bounce}). The insert shows the detailed evolution  around the bounce point. }
\label{Fig:bg}
\end{figure}

\begin{figure}[hb]
\centering
\includegraphics[scale=0.3]{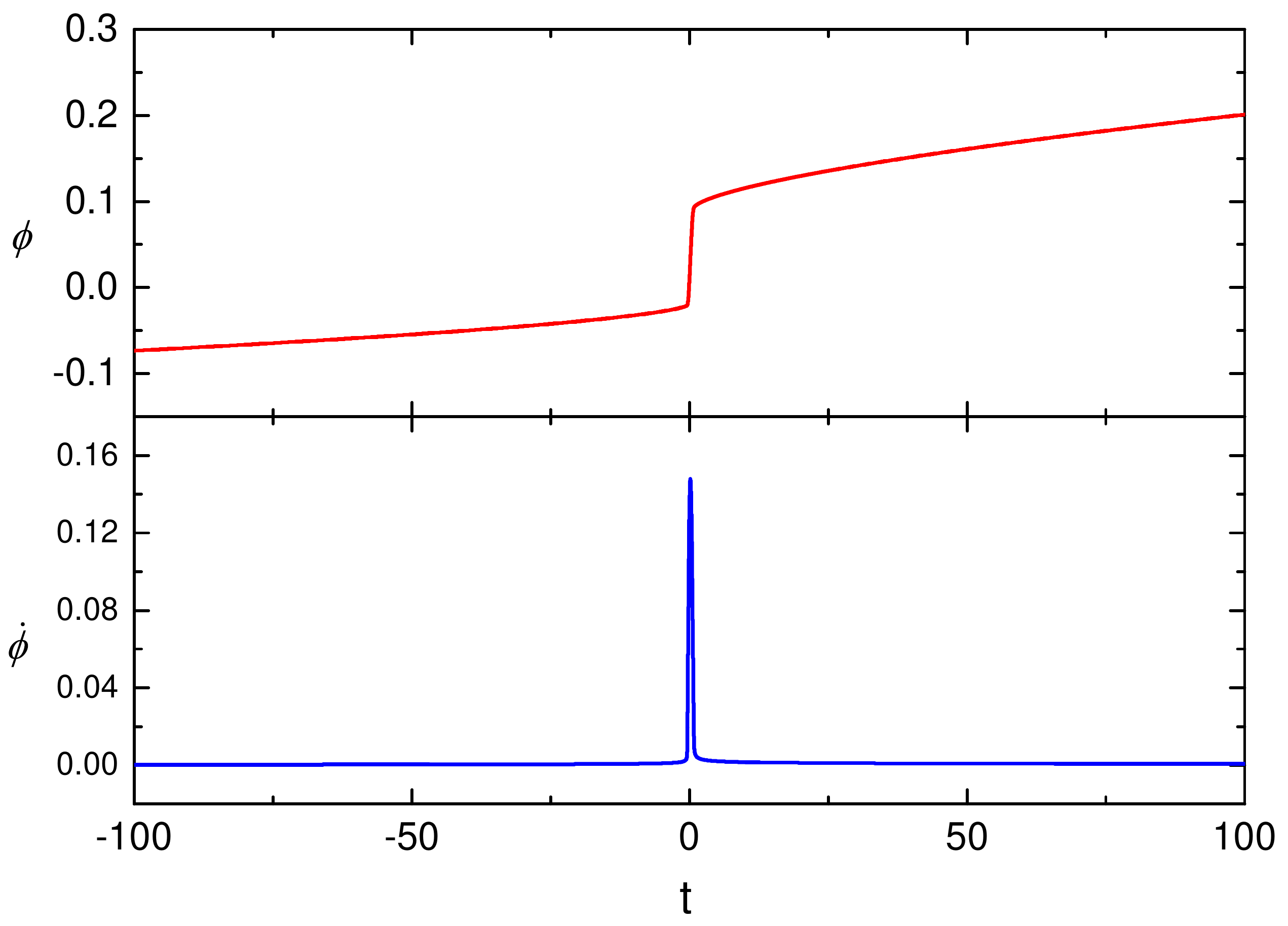}
\caption{Numerical evolution of the background scalar field $\phi$ and its time derivative $\dot{\phi}$ as functions of cosmic time when $f=f(X)$. }
\label{Fig:phi}
\end{figure}

By adopting $f=f(X)$, one can see from Equation \eqref{eq:fas} that $f_{a1} = f_{a2} = f_{a3} = 0$. Accordingly, the background equations of motion do not change when compared to the model proposed in \cite{Cai:2012va}. The numerical results for the evolution of the Hubble parameter $H$ and the equation-of-state parameter $w$ (which is defined as the ration between the pressure and energy density of the scalar field) are presented in Figure~\ref{Fig:bg}, and the numerical evolution of the background scalar field $\phi$ is provided in Figure~\ref{Fig:phi}. The zoomed-in views of the evolutions are also shown around the bounce point.

The parameters $z_s^2$ and $c_s^2$ from Equation \eqref{eq:zs2cs2} can be further simplified to be
\begin{align} 
&\frac{z_s^2}{2a^2} = 3 + \frac{2 \big[ \dot{\phi}^{2}( K_{X} - 2 Q_{\phi} ) + \dot{\phi}^{4} (K_{X X} - Q_{\phi X} ) - 6 H^{2} + 12 H Q_{X} \dot{\phi}^{3}+3 H \dot{\phi}^{5} Q_{X X} \big]} {\big( Q_{X} \dot{\phi}^{3}-2 H \big)^{2}}  ~, \nonumber \\
& (-\frac{z_s^2}{2a^2})c_s^2 = 1 + f + \frac{2}{a} \frac{d}{dt} \bigg[ \frac{a \big( f_{X} \dot{\phi}^{2} - f - 1 \big)}{2 H - Q_{X} \dot{\phi}^{3}} \bigg] ~.
\end{align}
Since the expression of $z_s^2$ is the same as in \cite{Cai:2012va}, the analyses on the positivity of $z^2$ remain the same. Therefore, the model at hand is free of ghost instability.

Let us now investigate the evolution of the $c_s^2$ and $c_T^2$ terms to examine whether the gradient instability can be safely solved in this specific example. For simplicity, we consider the form of $f(X)$ to be a polynomial function of $X$, namely $f(X)=\Sigma c_n X^n$. Additionally, the condition that, the propagation speed of tensor modes $c_T^2=1+f$ approaches $1$ in the late universe where $X \rightarrow 0$, can automatically yield the constraint $c_0=0$.

\begin{figure}[bh]
\centering
\includegraphics[width=0.45 \linewidth]{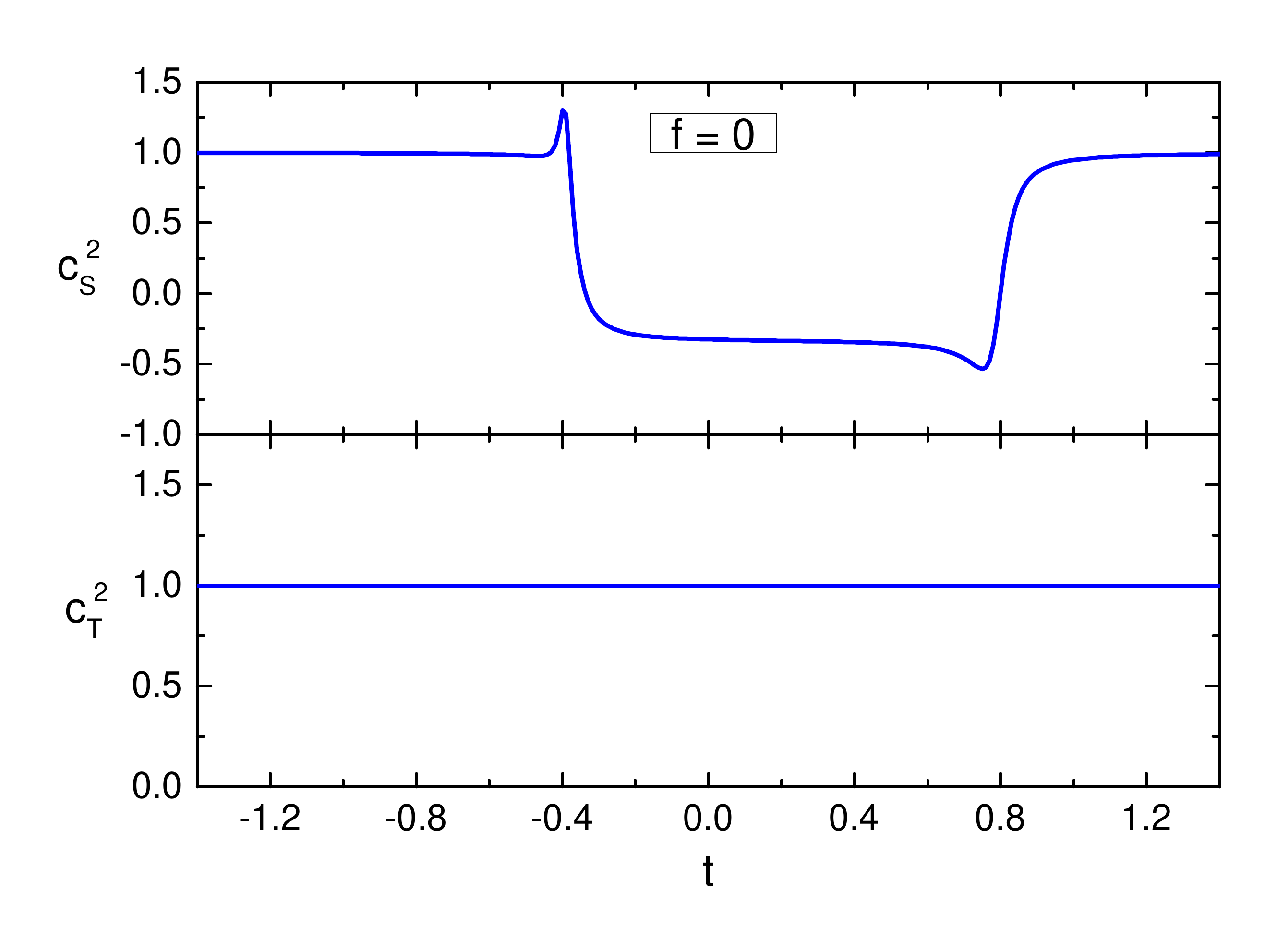}
\includegraphics[width=0.45 \linewidth]{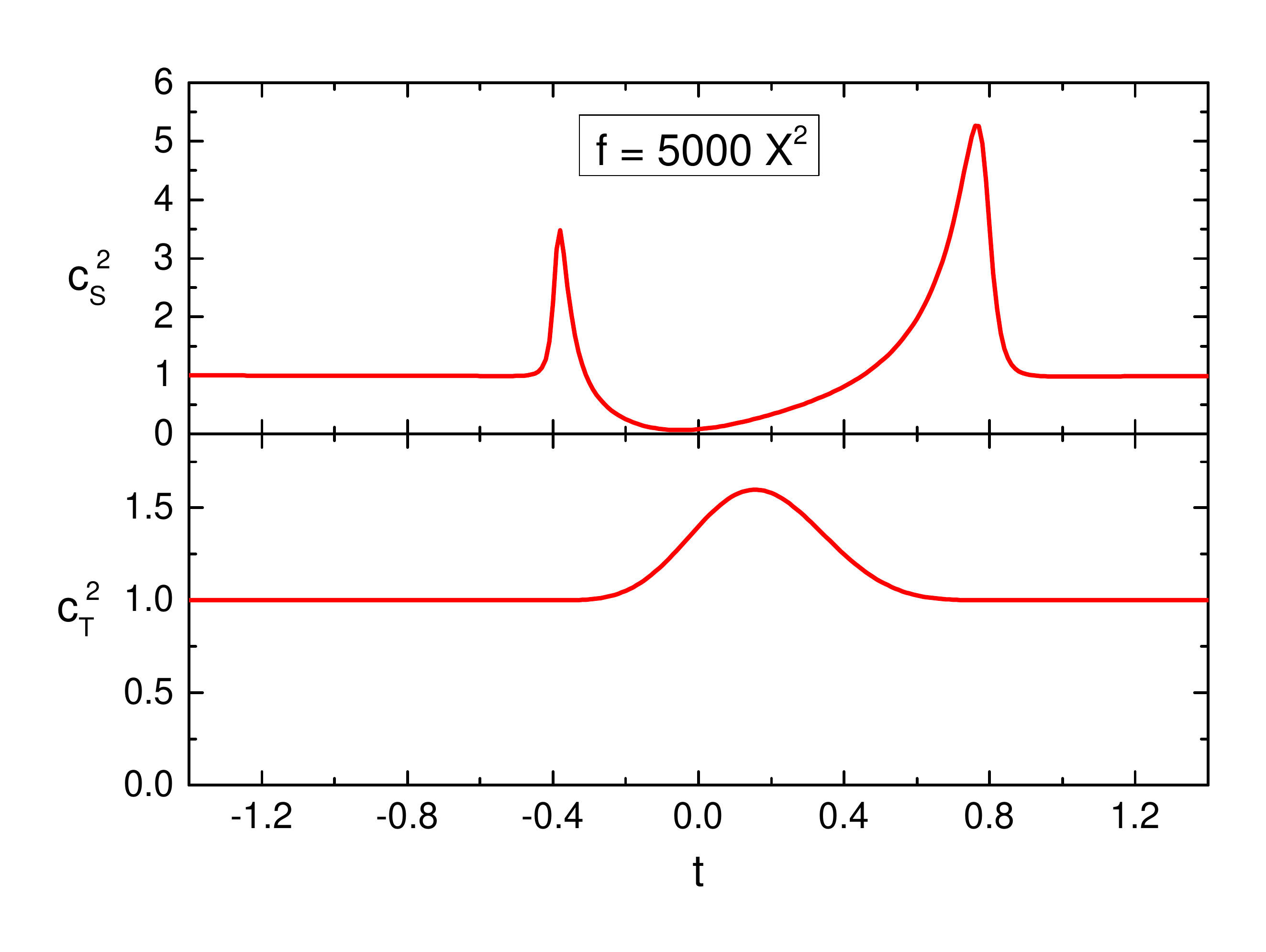}
\includegraphics[width=0.45 \linewidth]{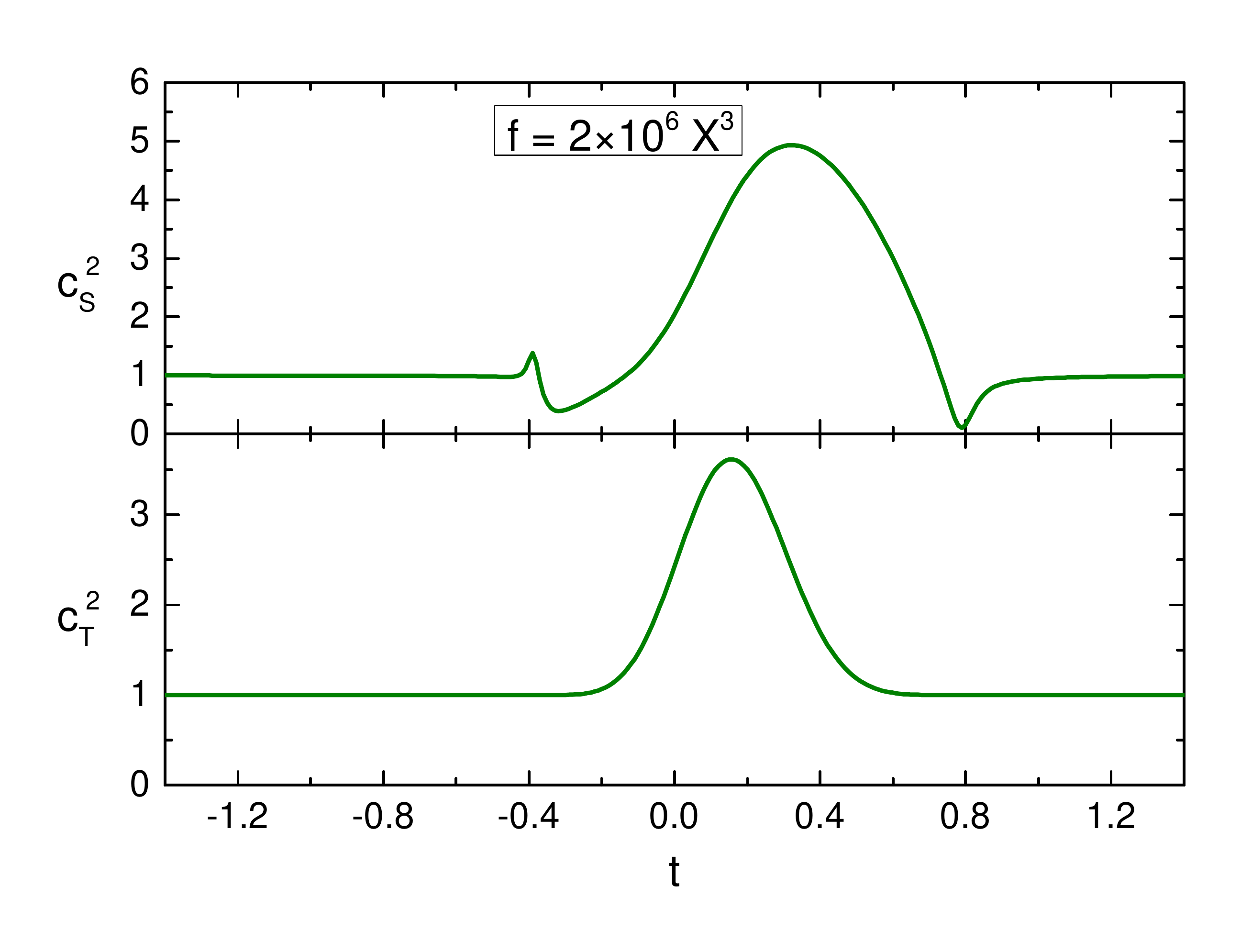}
\includegraphics[width=0.45 \linewidth]{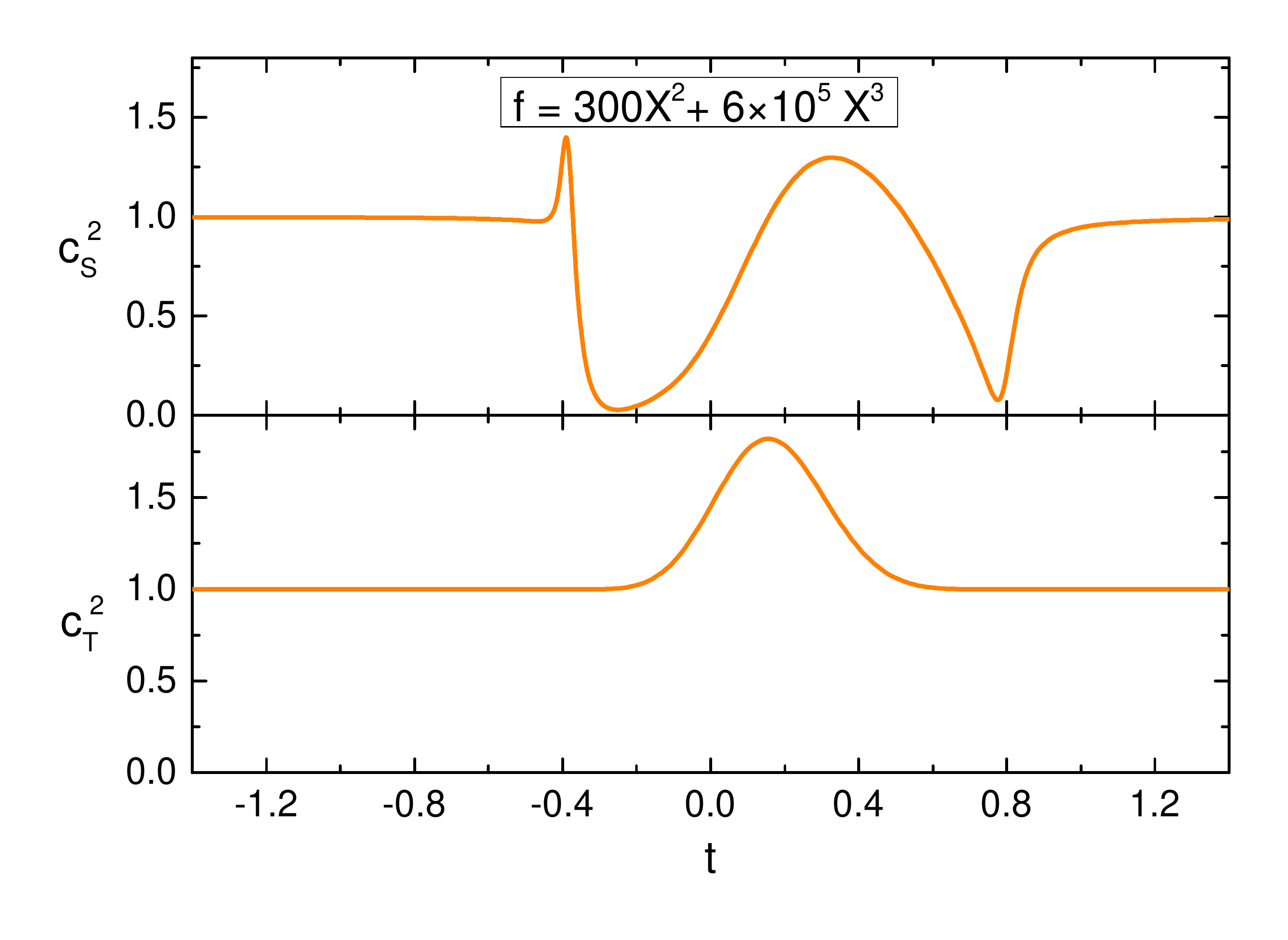}
\caption{Numerical evolution of the propagation speed squared, $c_s^2$ of scalar perturbations and $c_T^2$ of tensor perturbations, as functions of the cosmic time $t$ near the bounce point, with the DHOST function being $f = 0$ (top left), $f = 5000 X^2$ (top right), $f = 2 \times 10^6 X^3$ (bottom left) and $f = 300 X^2 + 6\times10^5 X^3$ (bottom right), respectively. The values of model parameters are provided in \eqref{eq:bgpara}. }
\label{fig:csctmodel1}
\end{figure}

With the parameters chosen in \eqref{eq:bgpara}, we can numerically solve $c_s^2$ and $c_T^2$ along with the background evolution. The results with different $f(X)$ forms are displayed in Figure \ref{fig:csctmodel1}. Note that, the function $f(X)$ is almost $0$ outside the bounce phase since $X \simeq 0$ according to Figure \ref{Fig:phi}, and hence the DHOST term yield little effect on $c_s^2$ and $c_T^2$ when it is away from the bounce. 
In this case, the dynamics of the perturbation will be similar to the case as analyzed in \cite{Cai:2012va}.
Thus, we only plot the behaviour of $c_s^2$ and $c_T^2$ near the bounce point (i.e. in the neighbourhood of $t=0$), for convenience.

The forms of $f(X)$ examined are:
\begin{enumerate}

\item $f=0$. In this special case our model can exactly reduce back to the original bounce model merely based on Horndeski theory as in \cite{Cai:2012va}, where there is no DHOST effect. As one can see in the left-top panel of Figure \ref{fig:csctmodel1}, we have $c_s^2<0$ lasting for a short period around the bounce point. This indicates exactly the gradient instability for nonsingular Horndeski bounces.

\item $f=c_1 X^2$. In this case we again find that the criterion of $c_s^2>0$ and $c_T^2 > 0$ can be always satisfied if we take a sufficiently large value of $c_2$ (for the specific parameter values that we chose for the background evolution \eqref{eq:bgpara} we obtain $c_2 \gtrsim 5000$). Additionally, for this example we also acquire $c_s^2 > 1$ and $c_T^2 > 1$ for a short interval.  

\item  $f=c_2 X^3$. In this case we find that the criterion of $c_s^2>0$ and $c_T^2 > 0$ can be always satisfied if $c_3>2*10^6$ with the background model parameters of \eqref{eq:bgpara}. Furthermore, there also exists a short period for $c_s^2 > 1$ and $c_T^2 > 1$ during the bounce phase.

\item $f = c_1 X^2 + c_2 X^3$. Having the above examples in mind, we now consider this combined model in order to fulfill the requirement of $c_s^2 > 0$ and $c_T^2 > 0$, and additionally to weaken the $c_s^2 > 1$ and $c_T^2 > 1$ phase around the bounce. As we observe in Figure \ref{fig:csctmodel1}, this can indeed be obtained.

\end{enumerate}

\subsection{Case 2: $f=f(\phi)$}
\label{sec:fphi}

In this subsection we further consider the case with $f=f(\phi)$ in order to show how generally the DHOST terms can cure the gradient instability in nonsingular bounce cosmologies. In this case the background equations of motion are apparently affected by the DHOST term. Nevertheless, by constructing explicit examples we find that nonsingular bounces without ghost and gradient instabilities can still be obtained.

\begin{figure}[htp]
\centering
\includegraphics[width=0.5 \linewidth]{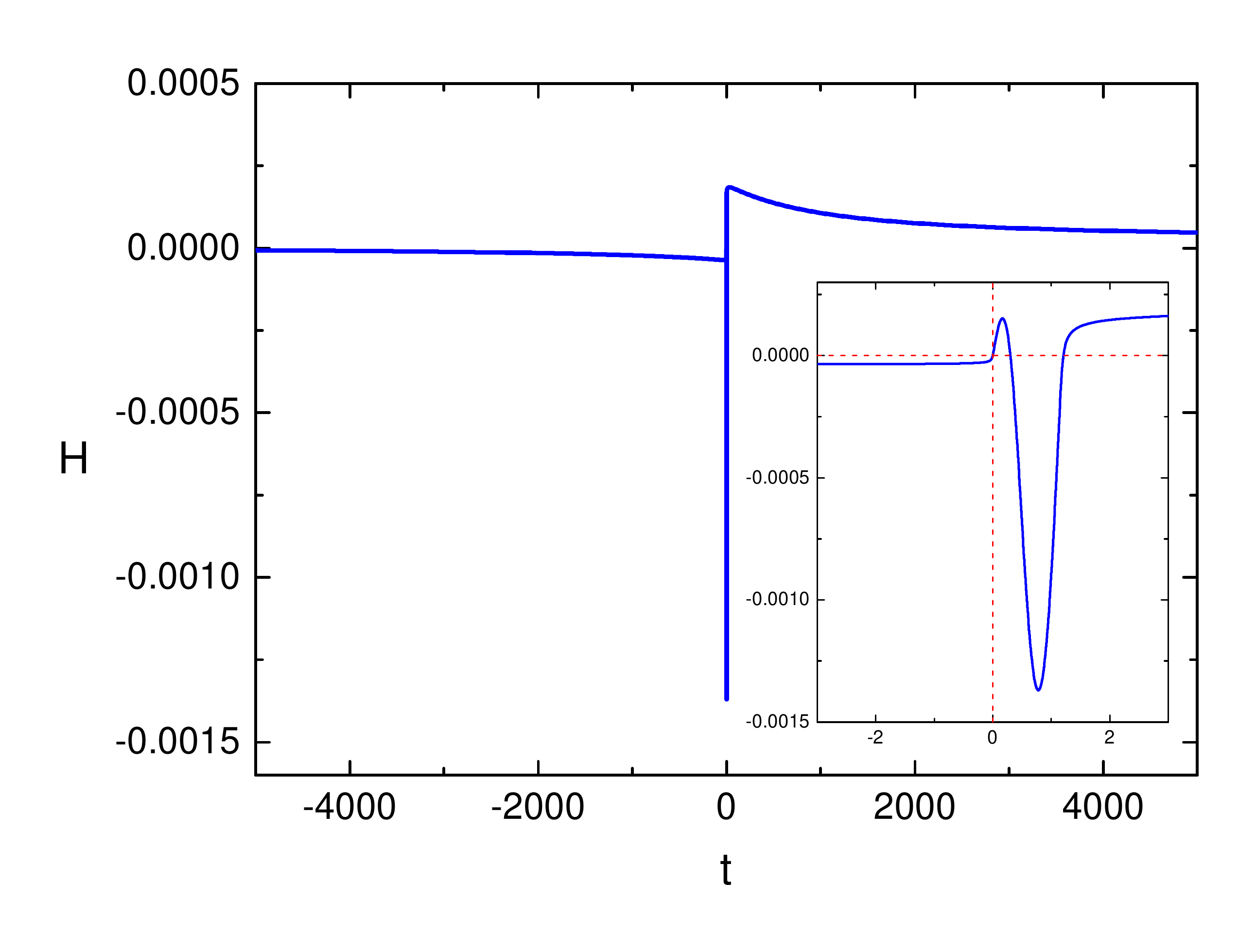}
\caption{Numerical evolution of the Hubble parameter $H$ of the background as a function of the cosmic time $t$, with the DHOST function being $f = e^{-0.1\phi^2}(1-e^{-0.2\phi^2})$. The values of model parameters are provided in \eqref{eq:bgpara}.
}
\label{fig:Hmodel2}
\end{figure}

We will examine one successful example $f = e^{-0.1\phi^2}(1-e^{-0.2\phi^2})$ in this section. We first draw the numerical evolution of the Hubble paramete $H$ to show that a nonsingular bounce can happen in Figure \ref{fig:Hmodel2}. Note that the bounces may happen multiple times, which is different from previous models.

\begin{figure}[bhtp]
\centering
\includegraphics[width=0.5 \linewidth]{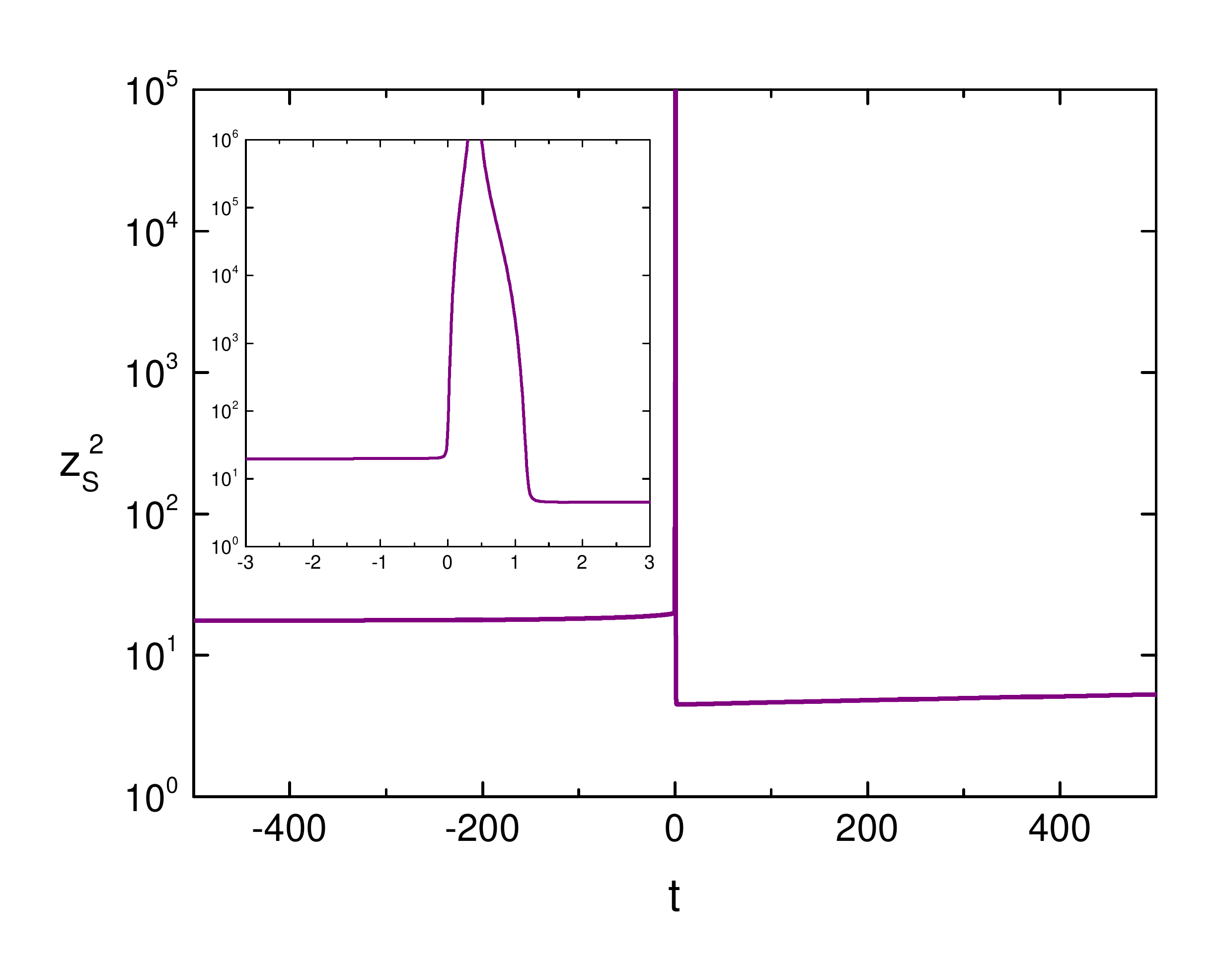}
\caption{Numerical evolution of $z_s^2$ as a function of cosmic time $t$, with the DHOST function being $f = e^{-0.1\phi^2}(1-e^{-0.2\phi^2})$. The values of model parameters are provided in \eqref{eq:bgpara}.
}
\label{fig:zs2fphi}
\end{figure}

We begin our perturbation analysis with the ghost issue. Since $z_T^2 = a^2/8$ is always positive, we only need to plot $z_s^2$ to show there is no ghost instability. The numerical evolution of $z_s^2$ is plotted in Figure \ref{fig:zs2fphi}. We can see from the numerical results that $z_s^2$ is always positive, hence the ghost problem is absent in our model.

\begin{figure}[bht]
	\centering
	\includegraphics[width=0.45 \linewidth]{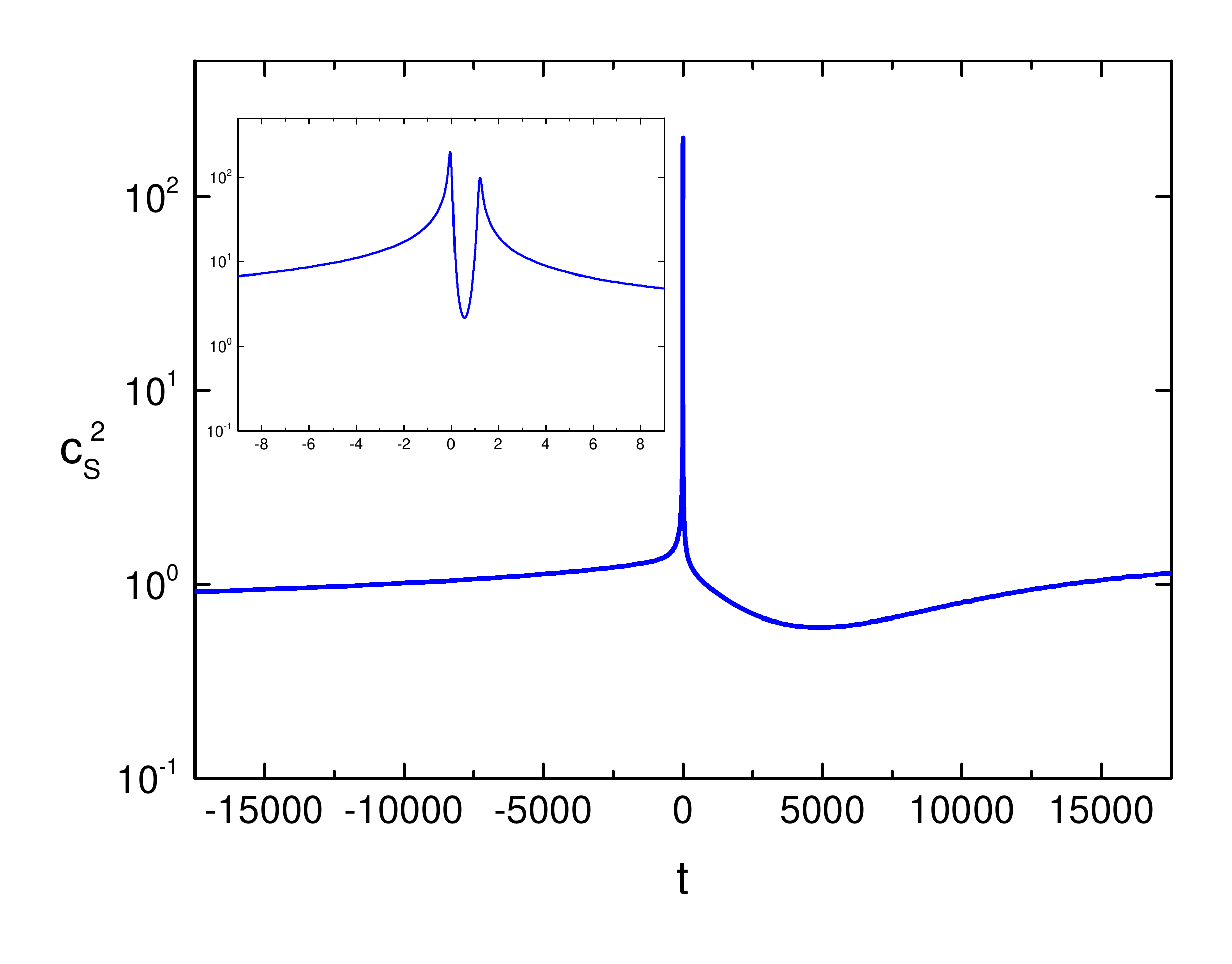}
	\includegraphics[width=0.45 \linewidth]{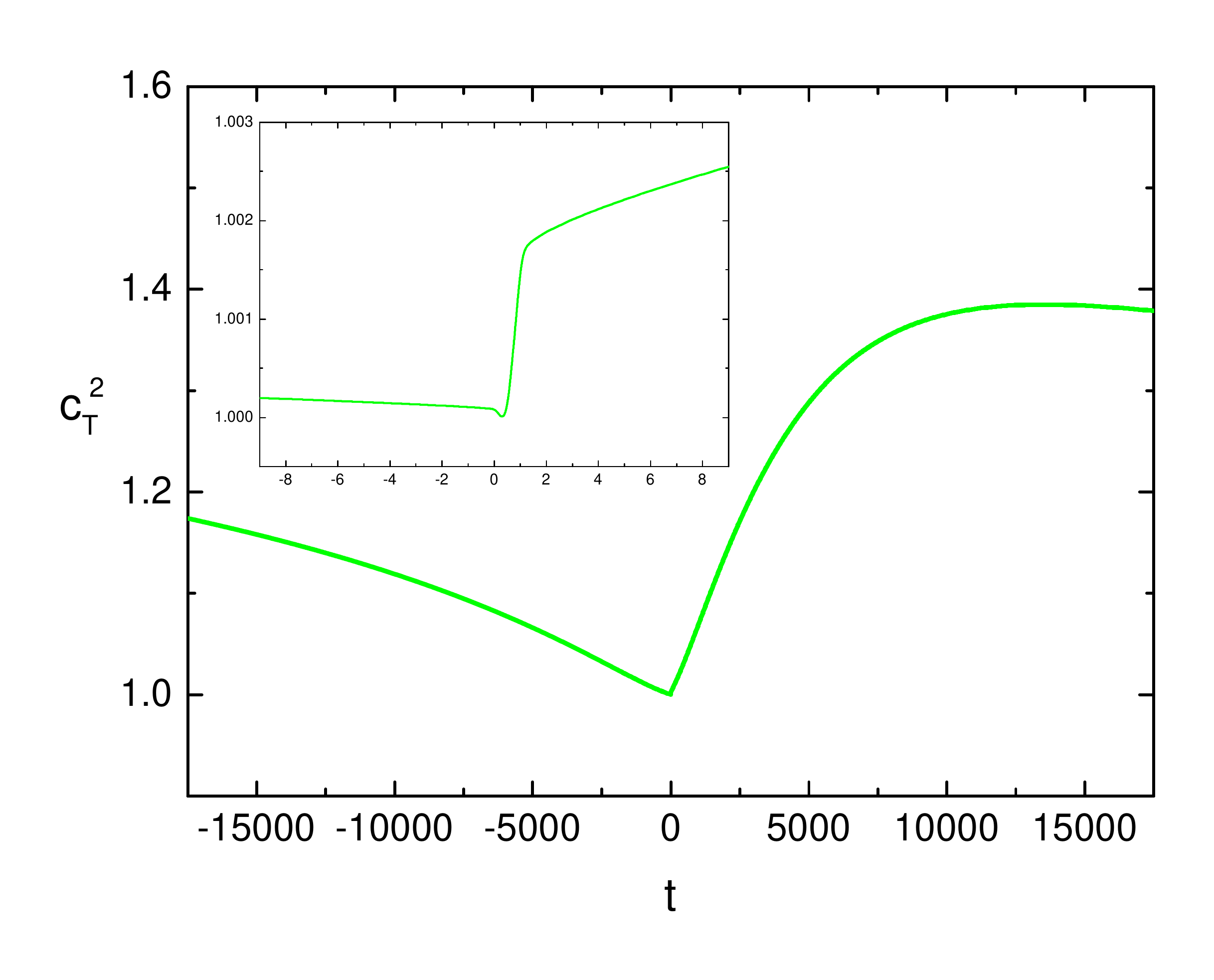}
	\caption{
		\it{Evolution of the sound speed squared $c_s^2$ and $c_T^2$ with the DHOST function being $f = e^{-0.1\phi^2}(1-e^{-0.2\phi^2})$.}
		\label{fig:csctmodel2}
	}
\end{figure}

Then we come to the propagation speed squared $c_s^2$ and $c_T^2$. We provide the numerical evolution of $c_s^2$ and $c_T^2$ in Figure \ref{fig:csctmodel2}. By explicitly showing their behaviour both globally and locally near the bounce, one can read that $c_s^2$ and $c_T^2$ are always positive, and therefore the gradient instability can be fully solved. 

\section{Super-luminality}
\label{sec:superluminality}
In the above explicit scenarios we showed that the DHOST corrections to the Horndeski bounces can remove the ghost and Laplacian instabilities by making $c_s^2$ to be larger than $0$ throughout the whole cosmological evolution. Nevertheless, as we have seen, there exist a short period around the bounce phase where the value of sound speed squared is larger than unity, i.e. $c_s^2 > 1$. This arises from the fact that the appearance of the DHOST terms can modify the dispersion relation of primordial perturbations and uplift the value of $c_s^2$ with added operators. Actually this is the mechanism that increases $c_s^2$ and allows us to avoid the $c_s^2 < 0$ regime. Although we should try to make a subtle fine-tuning in order to construct a model in which the aforementioned uplift of $c_s^2$ could simultaneously remove the $c_s^2 < 0$ regime but without entering into the $c_s^2 > 1$ region, there is no need to do that since in our case super-luminality is not problematic since it does not necessarily correspond to acausality \cite{ Bruneton:2006gf, Babichev:2007dw, Kang:2007vs, Deffayet:2010qz, Dobre:2017pnt}. In particular, super-luminality is known to be a general feature of large classes of Horndeski/Galileon theories, however, it does not imply pathologies or acausality necessarily, since superluminal propagation around specific solutions might imply simply that the theory cannot be UltraViolet-completed by a weekly coupled string theory or a Lorentz-invariant, local Quantum Field Theory \cite{Adams:2006sv}. In summary, in the scenarios of DHOST bounces examined in the present work we conclude that super-luminal propagation, although it could be removed by fine-tuning, is not problematic.

\section{Conclusion and discussions} \label{conclusion}

In this work, we have presented a new class of nonsingular bounce cosmology free from pathologies and instabilities, based on the DHOST theories. In this type of scenarios, the gradient instability that widely exists in nonsingular bounce cosmologies in the framework of scalar-tensor and Horndeski/Galileon theories is removed by the effects of the DHOST terms due to the modification that they later bring about to the dispersion relation of primordial perturbations. In this regard, the DHOST bounce cosmology can serve as a possibly healthy paradigm of the very early universe alternatively to inflation.

In the concrete realizations that we investigated, there can be at least two categories of bounce models, depending on the whether the DHOST function $f$ is a function of the scalar field $\phi$ or its kinetic term $X$. According to our detailed analysis, both types of models can avoid ghost as well as gradient instabilities throughout the whole background evolution. 

The above behaviour is obtained due to the fact that although the DHOST correction terms do not alter the background evolution, they do modify the dispersion relation of scalar perturbations, leading the squared sound speed to be uplifted and thus moving away from the $c_s^2 < 0$ regime that was plaguing the standard  Horndeski/Galileon  bounces. This uplifting may however lead $c_s^2$ to obtain superluminal values during sort periods of the evolution, nevertheless as we discussed this superluminality is not problematic since it does not correspond to acausality, and thus it does not need to get removed by fine-tuning. Finally, concerning  the  propagation speed of gravitational waves, we saw that although this is altered by the effect of the DHOST correction terms, one can suitably choose the function $f$ in order to obtain a late-time speed equal to the light speed and thus easily bypass the current gravitational wave constraints.

The  present study has illustrated the theoretical possibility of a healthy nonsingular bounce in the framework of DHOST theories. The construction of more  realistic scenarios should also incorporate confrontation with various observational constraints, including the high precision CMB measurement of primordial power spectrum, the tension between the tensor-to-scalar ratio and primordial non-gaussianities, as well as particle production and reheating process required to smoothly connect the observed thermal big bang expansion. All these topics deserve to be studied in a separate follow-up project.

\section*{Acknowledgement}
We are grateful to R. Brandenberger, D. Easson, X. Gao, A. Marciano, J. Quintin, T. Qiu, M. Sasaki, A. Vikman, D. G. Wang, Y. Wang and M. Yamaguchi for long-termed discussions and valuable comments. 
AI thanks the support of CAS-TWAS president fellowship for the PhD program. 
YFC is supported in part by the NSFC (Nos. 11722327, 11653002, 11961131007, 11421303), by the CAST-YESS (2016QNRC001), by the National Youth Thousand Talents Program of China, and by the Fundamental Research Funds for Central Universities.
YZ is supported in part by the NSFC (No. 11847239), and by the CAS Key Laboratory for Researches in Galaxies and Cosmology (No. 18010203). 
ENS is supported partly by the USTC fellowship for international visiting professors. This work is partially based upon work from COST Action ``Cosmology and Astrophysics Network for Theoretical Advances and Training Actions'', supported by COST (European Cooperation in Science and Technology). 
All numerics were operated on the computer clusters {\it LINDA}  \& {\it JUDY} in the particle cosmology group at USTC.

\appendix

\section{Quadratic DHOST Theory}\label{appen:dhost}

In order to evade Ostrogradski instabilities the Hessian matrix of the Lagrangian \eqref{dhost32}, using (\ref{extraterms}), should be degenerate. This imposes constraints on the choices of $a_i$'s and $b_i$'s. Following \cite{BenAchour:2016fzp}, all possible types of DHOST theories that are purely quadratic (namely with $f_3 = b_i = 0$) are listed as follows.

\begin{enumerate}

\item Minimally coupled theories. This category contains three cases. \\
(a) $^{(2)}M$\textendash $I$: There are three free functions $a_1$,$a_2$ and $a_3$, whose constraints are
\begin{equation} 
 a_4 = \frac{a_1}{X} ~,~~ a_5 = \frac{a_1 (a_1 +2a_2) +2a_1a_3 X  +3a_3^2 X^2}{4 (a_1+3a_2) X^2} ~,~~ a_2 \neq -\frac{a_1}{3} ~. \nonumber
\end{equation}
(b) $^{(2)} M$\textendash $II$: There are three free functions $a_1$, $a_4$, $a_5$, which are constrained by $a_2 = -\frac{a_1}{3}$ and $a_3=-\frac{a_1}{3X}$. \\
(c) $^{(2)} M$\textendash$III$: There  are four free functions $a_2$, $a_3$, $a_4$, $a_5$ and the unique condition $a_1=0$. There is only one scalar degree of freedom in this case. 

\item Non-minimally coupled theories. This category contains four subcases. \\
(a) $^{(2)} N$\textendash$I$: There are three free functions $f_2$,$a_1$ and $a_3$,  with $a_2 = -a_1 \neq \frac{f_2}{2X}$ and
\begin{align} 
 a_4 = & \frac{1}{8(f_2 + 2a_1X)^2}  \bigg\{ 4f_2 \Big[3(a_1 + f_{2X})^2-2a_3f_2 \Big] - 4 a_3 X^2 \big( -8 a_1 f_{2X}+ a_3 f_2 \big) \nonumber \\
 &
 \ \ \ \  \ \ \ \  \ \ \ \  \ \ \ \  \ \ \  \ \  - 8 X \big( 3 a_1 a_3 f_2 - 8 a_1^2 f_{2X} - 4 a_1 f_{2X}^2 - 4 a_1^3 - a_3 f_2 f_{2X} \big) \bigg\} ~, \nonumber \\
 a_5 = & \frac{1}{2 (f_2 + 2 a_1 X)^2} \big( a_1 + a_3 X + f_{2X} \big) \big[ a_1 (a_1 - 3 a_3 X + f_{2X}) - 2 a_3 f_2 \big] ~. \nonumber
\end{align}
(b) $^{(2)} N$\textendash$II$: There are three free functions $f_2$, $a_4$, $a_5$, together with $a_2 = -a_1 = \frac{f_2}{2X}$ and $a_3 = \frac{f_2 - 2 X f_{2X}}{2 X^2}$. \\
(c) $^{(2)} N$\textendash$III$: There are three free functions $f_2$, $a_1$, $a_2$, and the corresponding constraints are given by
\begin{align} 
 & a_1 + a_2 \neq 0 ~,~~ a_1 \neq -\frac{f_2}{2X} ~, \nonumber \\
 & a_3 = -\frac{2f_{2X}}{f_2} \big( a_1 + 3 a_2 \big) + \frac{( a_1 + 4a_2 + f_{2X} )}{X} -  \frac{f_2}{X^2} ~, \nonumber \\
 & a_4 = \frac{2f_{2X}^2}{f_2} + \frac{a_1 - f_{2X}}{X} + \frac{f_2}{2X^2} ~, \nonumber \\
 & a_5 = -\frac{1}{4f_2^2X^3} \Big[4f_2^3 - 2f_2^2X (3a_1 + 8a_2 + 6f_{2X}) + 8 f_2f_{2X}X^2(f_{2X} + 2a_1 + 6a_2) \nonumber \\ 
 & \ \ \ \  \ \ \ \  \ \ \ \  \ \ \ \  \ \ \ -12 f_{2X}^2 X^3 (a_1+3a_2) \Big] ~. \nonumber
\end{align}
	
\end{enumerate}

Note that since the conventions in \cite{BenAchour:2016fzp} are $X \equiv \nabla_{\mu} \phi \nabla^{\mu}\phi$ and the metric signature is $(-,+,+,+)$, the $X$ in \cite{BenAchour:2016fzp} should be replaced by $-2X$ when converting the convention into the present article.

Finally, we mention that the pure cubic  theory, namely with $f_2 = a_i = 0$, and the merging between cubic theory and quadratic theory, are much more complicated than the quadratic theory above, and since are not needed for the purpose of our analysis we  will not introduce them here.

\section{Quadratic action of cosmological perturbations}
\label{appen:perturbations}

In this Appendix we provide the quadratic action of cosmological perturbations of the above quadratic background action. To derive the quadratic action we decompose the metric $g_{\mu\nu}=n_{\mu}n_{\nu}-h_{\mu\nu}$, and the lapse function and shift vector are written as
\begin{align}
 N=1+\alpha ~,~ ~ N_{i}=\partial_{i} \sigma ~. 
\end{align}
The extrinsic curvature and acceleration are given by
\begin{align}
 K_{\mu\nu} &= \frac{1}{2} \mathcal{L}_n h_{\mu\nu} = \frac{1}{2} (\dot{h}_{ij}-N_{i|j}-N_{j|i}) ~, \nonumber \\
 a^{\mu} &= n^{\nu} \nabla_{\nu} n^{\mu} = h^{\mu i} \mathcal{D}_i \ln N ~,~ ~ a_\mu = h_{\mu \nu} a^\nu = \mathcal{D}_\mu \ln N ~. \nonumber
\end{align}
The Ricci scalar can be expressed as
\begin{align}
 -R = \mathcal{R} + K_{ij} K^{ij} - K^2 + 2 \nabla_\mu ( n^\mu K - a^\mu ) ~, \nonumber
\end{align}
where $\mathcal{R}$ is the Ricci scalar of the spatial coordinates. Afterwards, we apply the unitary gauge: 
\begin{align}
 \delta\phi=0 ~,~ ~ h_{i j}=a^{2} e^{2 \zeta} \delta_{i j} ~. 
\end{align}
Beside, under this gauge, we have the scalar quantity $\Box \phi = B K + W$, where
\begin{align}
 B = \phi_{,\mu} n^\mu = \frac{\dot{\phi}}{N} ~,~ ~ W = B_{,\mu} n^\mu = \frac{1}{N} \Big( \dot{B} + B \frac{N^{i} \partial_{i} N}{N} \Big) ~. \nonumber
\end{align}

Accordingly, the above basic quantities up to the second order become
\begin{align}
 & \mathcal{R} = -\frac{2}{a^{2}} \Big( \partial_{i} \zeta \partial^{i} \zeta + 2 \partial_{i} \partial^{i} \zeta - 4 \zeta \partial_{i} \partial^{i} \zeta \Big) ~, \nonumber \\
 & K_{ij} = a^{2} \delta_{i j} \Big[ H \big(1+2 \zeta-\alpha+\alpha^{2}-2 \zeta \alpha + 2 \zeta^{2} \big) + \dot{\zeta} \big(1 +2 \zeta -\alpha \big) \Big] \nonumber \\
 & \ \ \ \ \ \ \ \ \ \ \ \ \ \ \ \ - (1-\alpha) \partial_{i} \partial_{j} \sigma + \big( \partial_{i} \zeta \partial_{j} \sigma + \partial_{j} \zeta \partial_{i} \sigma - \partial^{k} \zeta \partial_{k} \sigma \delta_{i j} \big) ~, \nonumber \\ 
 & K = 3 H \big( 1-\alpha+\alpha^{2} \big) + 3 \dot{\zeta}(1-\alpha) - (1-2 \zeta-\alpha) \Delta \sigma - \frac{1}{a^{2}} \partial_{i} \zeta \partial^{i} \sigma ~, \nonumber \\
 & \mathcal{S} = K_{j}^{i} K_{i}^{j} = 3 H^{2} \big( 1-2 \alpha+3 \alpha^{2} \big) + 6  H \dot{\zeta}(1-2 \alpha) + 3 \dot{\zeta}^{2} \nonumber \\
 & \ \ \ \ \ \ \ \ \ \ \ \ \ \ \ \  - 2 \big( H -2 H \alpha -2 H \zeta +\dot{\zeta} \big) \Delta \sigma + \frac{1}{a^{4}} \partial_{i} \partial_{j} \sigma \partial^{i} \partial^{j} \sigma - \frac{2 H}{a^{2}} \partial^{i} \zeta \partial_{i} \sigma ~, \nonumber \\
 & K^{2} = 9 H^{2} \big( 1-2 \alpha+3 \alpha^{2} \big) + 9 \dot{\zeta}^{2} + (\Delta \sigma)^{2} + 18 H \dot{\zeta}(1-2 \alpha) - 6 \dot{\zeta} \Delta \sigma \nonumber \\
 & \ \ \ \ \ \ \ \ \ \ \ \ \ \ \ \ -6 H \big( 1 - 2 \alpha - 2 \zeta \big) \Delta \sigma - \frac{6 H}{a^{2}} \partial_{i} \zeta \partial^{i} \sigma ~, \nonumber \\ 
 & B = \dot{\phi}( 1 - \alpha + \alpha^2) , \nonumber \\  
 & W = \big(1 - 2 \alpha + 3 \alpha^2 \big) \ddot{\phi} - (1-3\alpha) \dot{\alpha} \dot{\phi} + \frac{\dot{\phi}}{a^2} \partial_{i} \alpha \partial^{i} \sigma ~. 
\end{align}

Now, for any scalar function such as $P$, $Q$ and $f$, the expansion up to second order becomes
\begin{align}
 P = \overline{P} + \frac{1}{2} \dot{\phi}^{2} \alpha \big( 3 \alpha-2 \big) \overline{P_{X}} + \frac{1}{2} \dot{\phi}^{4} \alpha^{2} \overline{P_{X X}} ~.
\end{align}
Hence, the quadratic part for the scalar perturbations of the Einstein-Hilbert action is given by
\begin{align} 
\label{4sh}
 S_{E H}^{(2)} = & \int d t d^{3} x a^{3} \Big[ -3 H^{2} \alpha^{2} +9 H^{2} \alpha \zeta - \frac{27}{2} H^{2} \zeta^{2} -3 \dot{\zeta}^{2} +6 H \dot{\zeta} (\alpha -3 \zeta) \nonumber \\
 & \ \ \ \ \ \ \ \ \ \ \ \ \ \ \ -2 H \alpha \Delta \sigma + 2 \dot{\zeta} \Delta \sigma + \frac{1}{a^{2}} \big( \partial_{i} \zeta \partial^{i} \zeta-2 \alpha \partial_{i} \partial^{i} \zeta \big) \Big] ~.
\end{align}
Additionally, for the part involving $K$ and $Q$ terms in Equation \eqref{eq:bounceS}, the quadratic action is then expressed as 
\begin{align} \label{skq}
 S_{K Q}^{(2)}  = & \int d t d^{3} x a^{3} \Big\{ \Big[ 3 \zeta \big( \alpha+\frac{3}{2} \zeta \big) K+\frac{1}{2} \dot{\phi}^{2} \big( \alpha^{2} - 6 \alpha \zeta \big) K_{X} +\frac{1}{2} \dot{\phi}^{4} \alpha^{2} K_{X X} \Big] \nonumber \\
 & \ \ \ \ \ \ \ \ \ \ \ \ \ \ \ + \big( \ddot{\phi}+3 H \dot{\phi} \big) \Big[ \big( \alpha^{2}-3 \alpha  
\zeta+\frac{9}{2} \zeta^{2} \big) Q+\frac{1}{2} \dot{\phi}^{2} \big( 5 
\alpha^{2}-6 \alpha \zeta \big) Q_{X}+\frac{1}{2} \dot{\phi}^{4} \alpha^{2} Q_{X 
X} \Big] \nonumber \\
 & \ \ \ \ \ \ \ \ \ \ \ \ \ \ \ + Q \dot{\phi} \Big[ \frac{1}{a^{2}} \partial^{i} \sigma \partial_{i}(\alpha-\zeta)+\dot{\alpha} \big( 2 \alpha -3 \zeta \big) + 3 \dot{\zeta} \big( 3 \zeta - \alpha \big) + \big( \alpha -\zeta \big) \Delta \sigma \Big] \nonumber \\ 
 & \ \ \ \ \ \ \ \ \ \ \ \ \ \ \ + Q_{X} \dot{\phi}^{3} \alpha \big( \dot{\alpha}-3 \dot{\zeta}+\Delta \sigma \big) \Big\} ~. 
\end{align}

Therefore, the total quadratic action for the DHOST theory is give by
\begin{align}
 S_2  = \int dt d^3x a^3 \mathcal{L}^{(2)} ~,
\end{align}
with
\begin{align} 
 \mathcal{L}^{(2)} = & - 3\dot{\zeta}^2 + \alpha^2  \Big[ \frac{1}{2} \dot{\phi}^2 K_X + \frac{1}{2} \dot{\phi}^4 K_{XX} + \big( \ddot{\phi} + 3 H \dot{\phi} \big) \big( Q + \frac{5}{2}\dot{\phi}^2 Q_X + \frac{1}{2}\dot{\phi}^4 Q_{XX} \big) - 3H^2 \nonumber \\
 & \ \ \ \ \ \ \ \ \ \ \ \ \ \ \ - \frac{3}{2} H \dot{\phi} \big( 2 f_{\phi} + \dot{\phi}^4  f_{\phi XX} + 5 \dot{\phi}^2 f_{\phi X} \big) \Big] \nonumber \\ 
 & + 3 \alpha \zeta \Big[ K - \dot{\phi}^2 K_X - (\ddot{\phi} + 3 H \dot{\phi}) \big( Q + \dot{\phi}^2 Q_X \big) + 3 H^2 + 3 H \dot{\phi} \big( f_{\phi} + \dot{\phi}^2 f_{\phi X} \big) \Big] \nonumber \\ 
 & + \frac{9}{2} \zeta^2 \Big[ K + \big( \ddot{\phi} + 3H \dot{\phi} \big) Q - 3 H^2 - 3 H f_{\phi} \dot{\phi} \Big] + \dot{\alpha} \dot{\phi} \Big[ \big( 2Q + Q_X \dot{\phi}^2 \big) \alpha - 3 Q \zeta \Big] \nonumber \\ 
 & + 3 \dot{\zeta} \Big\{ 3 \big[ \dot{\phi} (Q - f_{\phi} ) - 2H \big] \zeta - \alpha \big[ \dot{\phi} (Q - f_{\phi}) + \dot{\phi}^3 ( Q_X - f_{\phi X}) - 2H \big] \Big\} + \frac{1+f}{a^2} \partial_i \zeta \partial^i \zeta \nonumber \\ 
 & + \frac{2}{a^2} \big( f_X \dot{\phi}^2 - f - 1 \big) \alpha \partial_i \partial^i \zeta + 2 \dot{\zeta} \Delta \sigma + \big[ \dot{\phi}^3 (Q_X - f_{\phi X}) - f_{\phi}\dot{\phi}  - 2H \big] \alpha \Delta \sigma ~.
\end{align}
Variation with respect to $\sigma$ leads to  the constraint 
\begin{align}
 \alpha = \frac{2 \dot{\zeta}}{2 H - Q_{X} \dot{\phi}^{3} + f_{\phi} \dot{\phi} + f_{\phi X} \dot{\phi}^{3}} ~.
\end{align}

Using the above constraint, one can derive the simplified action of scalar modes at second order as: 
\begin{align}
 S_2 = \int d \tau d^{3} x \frac{z_s^{2}}{2} \Big[ \zeta^{\prime 2} - c_s^2 \big( \partial_{i} \zeta \big)^2 \Big] ~,
\end{align}
and where
\begin{align}
 &\frac{z_s^2}{2a^2} = 3 + 2 \Big[ (Q_{X}-f_{\phi X}) \dot{\phi}^{3} - f_{\phi} \dot{\phi} - 2 H \Big]^{-2} ~ \Big[ \dot{\phi}^{2} (K_{X} -2 Q_{\phi} ) + \dot{\phi}^{4} \big( K_{X X} -Q_{\phi X} \big) -6 H^{2} \nonumber \\
 & \ \ \ \ \ \ \ \ \ \ \ \ \ \ \ -3 H \dot{\phi} \Big( 2 f_{\phi}+\dot{\phi}^{4} f_{\phi X X} + 5 \dot{\phi}^{2} f_{\phi X} \Big) +12 H Q_{X} 
\dot{\phi}^{3} + 3 H \dot{\phi}^{5} Q_{X X} \Big] ~, \nonumber \\ 
&(-\frac{z_s^2}{2a^2})c_s^2 = 1+f +\frac{2}{a}\frac{d}{dt} \bigg[ \frac{a \big( f_{X} \dot{\phi}^{2}-f-1 \big) }{ 2 H-(Q_{X}-f_{\phi X}) \dot{\phi}^{3}+f_{\phi} \dot{\phi} } \bigg] ~. \nonumber
\end{align}

\bibliographystyle{unsrt}

\end{document}